\documentclass[
reprint,
 amsmath,
 amssymb,
 aip
 ]{revtex4-2}

\usepackage{graphicx}
\usepackage{mathtools}
\usepackage{xcolor}

\usepackage{color}
\usepackage{ulem}

\begin{document}

\title{Higher-order interactions for controlling time-delayed Kuramoto model}

\author{Narumi Fujii}
\affiliation{Department of Systems and Control Engineering, Institute of Science Tokyo, Tokyo 152-8552, Japan}
\affiliation{Department of Mathematics and Namur Institute for Complex Systems (naXys), University of Namur, 5000 Namur, Belgium}
\email{fujii.n.801e@m.isct.ac.jp}
\author{Martin Moriam\'e}
\affiliation{Department of Mathematics and Namur Institute for Complex Systems (naXys), University of Namur, 5000 Namur, Belgium}
\author{Maxime Lucas}
\affiliation{Department of Mathematics and Namur Institute for Complex Systems (naXys), University of Namur, 5000 Namur, Belgium}
\affiliation{Earth and Life Institute, Mycology, Catholic University of Louvain, B-1348 Louvain-la-Neuve, Belgium}
\author{Hiroya Nakao}
\affiliation{Department of Systems and Control Engineering, Institute of Science Tokyo, Tokyo 152-8552, Japan}
\affiliation{Research Center for Autonomous Systems Materialogy, Institute of Science Tokyo, Kanagawa 226-8501, Japan}
\author{Timoteo Carletti}
\affiliation{Department of Mathematics and Namur Institute for Complex Systems (naXys), University of Namur, 5000 Namur, Belgium}

\begin{abstract}

We propose a framework for controlling the collective dynamics of the time-delayed Kuramoto model
based on a delay-free, higher-order approximation of the delayed interactions.
By applying the Ott--Antonsen ansatz and the second-order averaging method to the resulting higher-order Kuramoto model,
we obtain a one-dimensional reduced equation for the order parameter dynamics. 
Numerical simulations demonstrate that the higher-order approximation predicts the dynamics of the original delayed system more accurately than the conventional pairwise approximation and enables the realization of bistability and intermediate synchronization states. 
Our results demonstrate the effectiveness of higher-order interpretations of time delays
for the control of oscillator networks with time-delayed interactions.

\end{abstract}

\maketitle
{\bf 
Synchronization, the emergence of coherent rhythms among interacting oscillators, plays important roles in biological and engineering systems. Real-world oscillator networks often involve time delays due to finite signal propagation speeds, but such delays are often neglected in mathematical modeling because they lead to infinite-dimensional delay-differential equations.
In this study, we propose a delay-free framework for controlling synchronization of
the time-delayed Kuramoto model using a higher-order approximation.
Numerical simulations demonstrate the effectiveness of the proposed approach, showing that the higher-order approximation provides more accurate predictions than the conventional pairwise approximation and enables the realization of various synchronization states.
}

\section{Introduction}
The collective synchronization in coupled oscillator systems has attracted considerable attention because of its fundamental importance in a wide range of engineering and biological applications~\cite{Winfree_1980, glass1988clocks,  Strogatz_1993, stefanovska1999physics, Pikovsky_2001, Kuramoto_2003, arenas2008synchronization, ermentrout2010mathematical, Golombek_2010, Karma_2013,  Dirk_2012,florian2013synchronization, Marco_2009}, including neuronal dynamics~\cite{ermentrout2010mathematical},
electric power grids~\cite{Dirk_2012}, and swarm robotics~\cite{Marco_2009}. 
In these systems, synchronization phenomena emerge through interactions among oscillators, and the resulting collective behavior influences the overall system performance and functionality. Therefore, developing effective methods for controlling synchronization is a central question in the study of complex dynamical networks~\cite{Moehlis_2006a, Moehlis_2006b, franci2012desynchronization, Zlotnik_2013, Jr-Shin_2016, Tanaka_2008, Kawamura_2008,sieber2014controlling,skardal2015control, Monga_2019, Takata_2021, Kato_2021, Ozawa_2021, Berman_2022, Petar_2023, Yawata_2024, Wilson_2024, Namura_2024a, Namura_2024b, martin2025hamiltonial}.

In many real-world systems, interactions among oscillators are not instantaneous but instead involve time delays arising from finite
speeds of signal transmission, communication latencies, or processing times~\cite{ermentrout1998fine, heil2001chaos, izhikevich2006polychronization, arenas2008synchronization}.
Such delays are known to significantly affect the collective dynamics of the system and can induce a variety of nontrivial behaviors that are absent in delay-free systems. Previous studies have reported that time delays can generate multiple synchronized states, hysteresis, and bistability between synchronized and incoherent states~\cite{schuster1989mutual,niebur1991collective, kuramotoc1991ollective,Yeung_1999,choi2000delay, ermentrout2010mathematical,strogatz2003delay,Juan2005kuramoto,mukesh2018synchronization,
kotani2012adjoint,kotani2020nonlinear,kyrychko2021dynamics,ameli2021timedelayed}. Time delays not only enrich the dynamics of coupled oscillator systems but also make their analysis and control substantially more challenging.

Motivated by their practical importance, many studies have investigated the effect of time delays in coupled oscillators~\cite{Juan2005kuramoto}. One typical approach is to approximate the effect of a small time delay as a phase lag in the coupling function up to the first order. This results in Sakaguchi--Kuramoto-type models without an explicit time delay~\cite{Sakaguchi_1986,bick2020understanding, pikovsky2024dynamics}, thereby avoiding the infinite-dimensional nature of delay differential equations~\cite{kori2001slow,lohe2015snchronization}. However, this first-order approximation cannot reproduce certain characteristic dynamical features of time-delayed oscillator systems. In particular, the standard Sakaguchi--Kuramoto model does not exhibit the bistability observed in time-delayed coupled oscillators. 

Recently, a new perspective on time-delayed coupled oscillator systems was introduced~\cite{bick2024timedelaysphaselags,fujii2025emergence,smirnov2026delayinertiatriadicinteractions}.
It has been shown that, beyond the first-order phase-lag approximation, time-delayed pairwise interactions can be interpreted as effective higher-order nonpairwise interactions, a framework that has attracted much attention in network science in recent years~\cite{battiston2020networks,carletti2020dynamical,bianconi2021higher,natphys,Dibakar_2022,bick2023higher,boccaletti2023structure,muolo2024turing,maxime2024deeper,millan2025topology,battiston2025collective}.
In particular, it was demonstrated that the bistability observed in time-delayed coupled oscillator networks can be successfully reproduced by using effective three-body interactions without delays~\cite{bick2024timedelaysphaselags,fujii2025emergence}.
This result shows that direct analysis of infinite-dimensional delayed systems can be avoided by representing the delays as finite-dimensional higher-order interactions.

In this work, based on this perspective, we propose a delay-free control framework for collective synchronization of the time-delayed Kuramoto model. Instead of explicitly incorporating time delays into the control design, the proposed framework captures the dynamical effects of delays through higher-order interactions,
enabling the design of tractable control strategies while preserving the essential synchronization properties induced by delays. 
We consider two control schemes, linear and nonlinear mean-field feedback. 
The linear feedback control can stabilize both the fully-synchronized and incoherent states, and can induce their bistability.
In the nonlinear feedback control, control parameters can be designed to stabilize arbitrary synchronization levels based on the reduced model.
In both cases, the resulting synchronization states are realized in the original time-delayed Kuramoto model
through numerical simulations.
Through this framework, we aim to provide a new methodology for analyzing and controlling delayed coupled oscillator systems from the viewpoint of higher-order network dynamics.

This paper is organized as follows.
First, we introduce the Kuramoto model with time-delayed pairwise coupling under a control input in Sec.~\ref{sec:TDKmodel}.
Next, we derive a delay-free model by higher-order approximation of time-delayed interactions
and use the Ott--Antonsen (OA) ansatz~\cite{Ott_Antonsen_2008}
and second-order averaging~\cite{verhulst2007averaging} to derive a one-dimensional 
equation for the order parameter in Sec.~\ref{sec:Avedyn}.
In Sec.~\ref{sec:feedbck}, we consider linear and nonlinear mean-field feedback control schemes and construct phase diagrams using the reduced equation, and Sec.~\ref{sec:numsim} presents numerical results for the original time-delayed Kuramoto model to validate our framework.
We also compare the higher-order approximation with a conventional pairwise approximation. Finally, Sec.~\ref{sec:conc} concludes the paper.

\section{Time-delayed Kuramoto model and higher-order approximation}
\label{sec:TDKmodel}

We consider a system of globally coupled $N$ phase oscillators with time-delayed pairwise interactions under a control input, given by the following set of delay-differential equations~\cite{Yeung_1999}:
\begin{align}
    \dot{\theta}_j(t)=\omega_j+\frac{\epsilon}{N}\sum_{\substack{k=1\\k\neq j}}^{N}\sin\left(\theta_k(t-\tau)-\theta_j(t)\right)+\epsilon\sin\left(\theta_j(t)\right)u(t),
    \label{eq:Kuramoto_time_delay_with_control}
\end{align}
for $j=1, ..., N$, where $\theta_j$ and $\omega_j$ are the phase and the natural frequency of $j$th oscillator.
In the coupling term, $\epsilon > 0$ denotes the coupling strength, which is assumed to be small, 
and $\tau > 0$ is a time delay in the transmission of the phase information. 
In the control term, we employ $\sin\left(\theta_j(t)\right)$ as 
the phase sensitivity function~\cite{Kuramoto_2003,Nakao_2016}, and $u(t)$ is the control input.
\par

In this study, we consider identical oscillators with $\omega_j \sim \delta(\omega-\omega_0)$.
To derive a macroscopic description using the OA ansatz, we temporarily assume that the natural frequencies follow the Lorentzian distribution $g(\omega)=\frac{\gamma/\pi}{(\omega-\omega_0)^2+\gamma^2}$,
where $\omega_0$ is the central frequency and $\gamma$ is the width parameter. The identical-oscillator case is then recovered by taking the limit $\gamma\rightarrow0^+$.

We introduce the complex order parameter defined as $z(t)=R(t)e^{i\Psi(t)}=1/N\sum_{j=1}^N e^{i\theta_j(t)}$, where $R$ represents the degree of synchronization and $\Psi$ denotes the collective phase.
Our control objective consists in achieving a desired degree of synchronization, i.e., $R(t)$ close enough to $1$, by using a control input $u(t)$.
However, designing $u(t)$ is challenging because Eq.~\eqref{eq:Kuramoto_time_delay_with_control} 
is an infinite-dimensional system due to the time delay. 
Therefore, in the following, we derive a reduced one-dimensional ordinary differential equation describing the time evolution of the order parameter $R$.

As shown in Ref.~\onlinecite{fujii2025emergence}, 
the time-delayed interactions in Eq.~\eqref{eq:Kuramoto_time_delay_with_control} can be 
approximated by higher-order interactions for small coupling strength $\epsilon$,
yielding the following delay-free Kuramoto model with higher-order interactions for the identical case:
\begin{widetext}
\begin{align}
    \dot{\theta}_j(t)
    &=\omega_0+\frac{\epsilon}{N}\sum_{\substack{k=1\\k\neq j}}^{N}\sin\left(\theta_k(t)-\theta_j(t)
    -\omega_0\tau\right)\notag\\
    &\quad+\frac{\epsilon^2\tau}{2N^2}\sum_{\substack{k=1\\k\neq j}}^{N}\sum_{\substack{l=1\\l\neq k}}^{N}
    \Big\{-\sin\left(\theta_l(t)-\theta_j(t)-2\omega_0\tau
    \right)+\sin\left(2\theta_k(t)-\theta_l(t)-\theta_j(t)\right)\Big\}+\epsilon\sin\left(\theta_j(t)\right)u(t),
    \label{eq:higher-order_model_with_control}
\end{align}
\end{widetext}
where terms $\mathcal{O}(\epsilon^3 \tau^2)$ are neglected.
The accuracy of this approximation depends on the smallness of $\epsilon \tau$.
Equation~\eqref{eq:higher-order_model_with_control} is an $N$-dimensional ordinary differential equations where the time delay appears as phase lags.
For simplicity, we do not assume delays in the control input and focus on delays in the interactions between oscillators.

\section{Averaged dynamics of the order parameter}
\label{sec:Avedyn}

Our next step is to apply the OA ansatz to derive a two-dimensional dynamical model for the complex order parameter $z$.
Following Ref.~\onlinecite{fujii2025emergence}, we first assume that the natural frequencies $\omega_j$ in Eq.~\eqref{eq:Kuramoto_time_delay_with_control} obey a Lorentzian distribution and derive the corresponding higher-order Kuramoto model.
We then consider the dynamics of the probability density function $P(\theta,~\omega,~t)$ 
of the oscillator phase $\theta$ and frequency $\omega$
in the limit $N \rightarrow \infty$, where the complex order parameter is
redefined as $z(t)=R(t)e^{i\Psi(t)}=\int_{-\infty}^{\infty}\int_{0}^{2\pi}e^{i\theta'}
P(\theta',~\omega',~t)g(\omega')d\theta'd\omega'$.
Applying the OA ansatz to $P(\theta,~\omega,~t)$ and subsequently taking the limit $\gamma \rightarrow 0^+$
to approximate the homogeneous case with $\omega_j\sim\delta(\omega-\omega_0)$, we obtain an equation for $z$ given by
\begin{widetext}
\begin{align}
    \dot{z}=
    \left\{i\omega_0+\frac{\epsilon}{2} e^{-i\omega_0 \tau}-\frac{\tau \epsilon^2}{4} e^{-2 i\omega_0 \tau}
    -\left(\frac{\epsilon}{2} e^{i\omega_0 \tau}-\frac{\tau \epsilon^2}{4} e^{2 i\omega_0 \tau}-\frac{\tau \epsilon^2}{4}\right)|{z}|^2 -\frac{\tau \epsilon^2}{4}|{z}|^4\right\} z+
    \frac{\epsilon}{2}u(t)\left(z^2-1\right).
    \label{eq:OA_with_control}
\end{align}
\end{widetext}
Note that the OA ansatz can also be applied directly to the time-delayed Kuramoto model~\eqref{eq:Kuramoto_time_delay_with_control}~\cite{Ott_Antonsen_2008, Ott_Antonsen_2009}. 
In this case, the derived equation is a delay differential equation for the complex order parameter $z$, and 
Eq.~\eqref{eq:OA_with_control} can be obtained by performing an expansion of $z(t-\tau)$
(See Ref.~\onlinecite{fujii2025emergence} for details).
Let us observe that the proposed approach allows us to explicitly reveal that the 
resulting system can be interpreted as a higher-order interaction approximation.

In this work, we employ a mean-field feedback control described by 
\begin{align}
    u(t)= D(R(t))\cos\Psi(t),
    \label{eq:control0}
\end{align}
where $D$ is a real function of the order parameter $R$.
From Eq.~(\ref{eq:OA_with_control}), the evolution equations for $R$ and $\Psi$ are given by
\begin{widetext}
{\small
\begin{subequations}
\begin{align}
    \dot{R}&=\left(\frac{\epsilon}{2}\cos(\omega_0 \tau)-\frac{\tau \epsilon^2}{4}\cos(2\omega_0\tau)\right)R-\left(\frac{\epsilon}{2}\cos(\omega_0 \tau)
    -\frac{\tau \epsilon^2}{4}-\frac{\tau \epsilon^2}{4}\cos(2\omega_0\tau)\right){R^3}-\frac{\tau \epsilon^2}{4}R^5
    +\frac{\epsilon}{2}D\left(R\right)\left(R^2-1\right)(\cos\Psi)^2,\label{eq:R_OA}\\
    \dot{\Psi}&=\omega_0 -\frac{\epsilon}{2}\sin(\omega_0 \tau)+\frac{\tau \epsilon^2}{4}\sin(2\omega_0\tau)
    -\left(\frac{\epsilon}{2}\sin(\omega_0 \tau)-\frac{\tau \epsilon^2}{4}\sin(2\omega_0\tau)\right){R^2}+\frac{\epsilon}{2R} D\left(R\right)(R^2+1)\sin\Psi\cos\Psi,\label{eq:Psi_OA}
\end{align}
\label{eq:OA_with_control_R_Psi}
\end{subequations}
}
\end{widetext}
respectively, where the equation for $\Psi$ is valid for $R \neq 0$.
Note that $R$ is a slow variable because the right-hand side is $O(\epsilon)$, while $\Psi$ evolves rapidly due to the $\omega_0=\mathcal{O}(1)$ term.

Our control objective is to achieve a desired degree of synchronization, quantified by $R$. 
Therefore, it is desirable to derive a one-dimensional equation for $R$
by averaging out the fast dynamics of $\Psi$.
To this end, we apply the second-order averaging method to Eq.~\eqref{eq:OA_with_control_R_Psi}.
Note that the second-order averaging is necessary because we need to achieve the accuracy of $O(\epsilon^2)$, at which the effect of the higher-order interactions appears.
We rewrite Eq.~\eqref{eq:OA_with_control_R_Psi} as
\begin{subequations}
\begin{align}
    \dot{R}&=\epsilon F^{(1)}\left(R, \Psi\right)+\epsilon^2F^{(2)}\left(R\right),\label{eq:R_dynamics_simple}\\
    \dot{\Psi}&=\omega_0+\epsilon G^{(1)}\left(R, \Psi\right)+\epsilon^2G^{(2)}\left(R\right),
\end{align}
\label{eq:OA_with_control_R_Psi_2}
\end{subequations}
where 
{\small
\begin{subequations}
\begin{align}
    &F^{(1)}\left(R, \Psi\right)=\frac{1-R^2}{2}\left(\cos(\omega_0 \tau)R
    -D\left(R\right)(\cos\Psi)^2\right),\\
    &G^{(1)}\left(R, \Psi\right)=-\frac{1+R^2}{2}\left(\sin(\omega_0 \tau)
    -\frac{ D\left(R\right)}{2R}\sin2\Psi\right),\\
    &F^{(2)}\left(R\right)=-\frac{\tau}{4}R\left(1-R^2\right)\left(\cos(2\omega_0\tau)-
    {R^2}\right),\\
    &G^{(2)}\left(R\right)=\frac{\tau}{4}\left(1+{R^2}\right)\sin(2\omega_0\tau).
\end{align}
\end{subequations}
}\noindent
We introduce a near-identity transformation from the original variables, $R, \Psi$, to the averaged ones, $\bar{R}$, $\bar{\Psi}$, described as
\begin{subequations}
\begin{align}
    R&=\bar{R}+\epsilon g^{(1)}\left(\bar{R}, \bar{\Psi}\right)+\epsilon^2g^{(2)}\left(\bar{R}, \bar{\Psi}\right),\label{eq:new_R}\\
    \Psi&=\bar{\Psi}+\epsilon l^{(1)}\left(\bar{R}, \bar{\Psi}\right)+\epsilon^2l^{(2)}\left(\bar{R}, \bar{\Psi}\right),\label{eq:new_Psi}
\end{align}
\label{eq:transformation}
\end{subequations}\noindent\hspace{-3mm}
where $g^{(1)},~g^{(2)},~l^{(1)},~l^{(2)}$ are determined such that
the evolution of the averaged variables $\bar{R}$ and $\bar{\Psi}$ depend only on the slow averaged variable $\bar{R}$:
\begin{subequations}
\begin{align}
    \dot{\bar{R}}&=\epsilon h^{(1)}\left(\bar{R}\right)+\epsilon^2 h^{(2)}\left(\bar{R}\right)\label{eq:averaging_R},\\
    \dot{\bar{\Psi}}&=\omega_0+\epsilon q^{(1)}\left(\bar{R}\right)+\epsilon^2 q^{(2)}\left(\bar{R}\right).\label{eq:averaging_Psi}
\end{align}\label{eq:averaging}
\end{subequations}\noindent\hspace{-3.5mm}
Then, Eq.~\eqref{eq:averaging_R} provides an averaged one-dimensional equation for the transformed $\bar{R}$.
By evaluating $g^{(j)},~l^{(j)},~h^{(j)}$ and $q^{(j)}$ for $j=1,~2$ 
(See Appendix~\ref{app:derive_one} for details), 
we obtain the averaged one-dimensional equation for $\bar{R}$ from Eq.~\eqref{eq:averaging_R} as
\begin{widetext}
\begin{align}
\dot{\bar{R}}
&=
(1-\bar{R}^2)
\left\{
\bar{R}
\left(
\frac{\epsilon}{2}\cos(\omega_0\tau)
-\frac{\tau\epsilon^2}{4}\cos(2\omega_0\tau)
+\frac{\tau\epsilon^2}{4}\bar{R}^2
\right)
-\frac{\epsilon D(\bar{R})}{4}
\right\}.
\label{eq:averaging_Rh}
\end{align}
\end{widetext}

\section{Feedback control}
\label{sec:feedbck}

In this section, we design the control input $u(t)$, given in the form of Eq.~\eqref{eq:control0},
based on the averaged one-dimensional equation for averaged $\bar{R}$ in Eq.~\eqref{eq:averaging_Rh}.
We consider two types of mean-field feedback control defined by using the original variables $R$ and $\Psi$. 
The first case is a linear mean-field feedback control~\cite{franci2011existence, franci2012desynchronization,Ozawa_2021} in $R$ given by
\begin{align}
    u(t)=CR(t) \cos\Psi(t),
    \label{eq:control1}
\end{align}
where $C=\mathcal{O}(1)$ is a control gain and $D\left(R\right)=CR$.
For the second case, we propose
\begin{align}
    u(t)=CR(t) \left(R(t)-E\right)\cos\Psi(t),
    \label{eq:control2}
\end{align}
which gives a nonlinear feedback in $R$ and corresponds to $D\left(R\right)=CR(R-E)$.
In this case, we have another control parameter $E=\mathcal{O}(1)$ in addition to the gain $C=\mathcal{O}(1)$.
As demonstrated later, the nonlinear mean-field feedback control enables stabilization of arbitrary synchronization levels, which cannot be achieved by the  linear mean-field feedback control.

For each case, we design the control input based on two types of reduced equations for comparison.  
The first is the averaged one-dimensional equation derived via the higher-order approximation, given by Eq.~\eqref{eq:averaging_Rh}. 
The second is the averaged one-dimensional equation derived from the conventional pairwise approximation, described by the Sakaguchi--Kuramoto model~\cite{Sakaguchi_1986}, 
in which $\mathcal{O}(\tau\epsilon^2)$ terms are neglected in Eq.~\eqref{eq:higher-order_model_with_control}, yielding
\begin{align}
    \dot{\theta}_j(t)
    =\omega_0+\frac{\epsilon}{N}\sum_{\substack{k=1\\k\neq j}}^{N}
    \sin\!\left(\theta_k(t)-\theta_j(t)-\omega_0\tau\right)
    +\epsilon\sin\left(\theta_j(t)\right)u(t),
    \label{eq:pairwise_model_with_control}
\end{align}
and the subsequent procedure is essentially the same as in the higher-order approximation.
Eventually, the corresponding averaged one-dimensional equation is obtained as
\begin{align}
\dot{\bar{R}}
&=
(1-\bar{R}^2)
\left\{
\frac{\epsilon}{2}\cos(\omega_0\tau)\bar{R}
-\frac{\epsilon D(\bar{R})}{4}
\right\}.
\label{eq:averaging_Rp}
\end{align}
Note that Eq.~\eqref{eq:averaging_Rp} is also derived from Eq.~\eqref{eq:averaging_Rh}
by neglecting $\mathcal{O}(\tau\epsilon^2)$ terms.
In the following, we refer to Eq.~\eqref{eq:averaging_Rh}
as the higher-order reduced equation, 
and Eq.~\eqref{eq:averaging_Rp} as the pairwise reduced equation.

\subsection{Case I: $u=C R \cos\Psi$}

In this subsection,
we prove that the incoherent state given by $\bar{R}=0$ and
the fully-synchronized state given by $\bar{R}=1$, solutions of Eqs.~\eqref{eq:averaging_Rh} and \eqref{eq:averaging_Rp},
are stabilized under the control input given by Eq.~\eqref{eq:control1}.

\subsubsection{Higher-order approximation}

\label{subsubsec:higher-order_control1}

From Eq.~\eqref{eq:averaging_Rh} and $D\left(\bar{R}\right)=C\bar{R}$, we obtain
the higher-order reduced equation
{\small
\begin{align}
    \dot{\bar{R}}&=\bar{R}(1-\bar{R}^2)
    \left\{\left(
    \frac{\epsilon}{2}\cos(\omega_0\tau)
    -\frac{\tau\epsilon^2}{4}\cos(2\omega_0\tau)
    +\frac{\tau\epsilon^2}{4}\bar{R}^2
    \right)
    -\frac{\epsilon C}{4}
    \right\}\notag\\
    &\eqqcolon f_{\rm h}^1\left(\bar{R}\right),
    \label{eq:higher_order_control1}
\end{align}
}\noindent
where we defined the right-hand side as $f_{\rm h}^1$.
From $f_{\rm h}^1\left(\bar{R}\right)=0$, the fixed points are $\bar{R}^*=0,~1$, and $\sqrt{-\frac{\alpha}{\beta}+\frac{\epsilon C}{4\beta}}$,
where $\alpha=\frac{\epsilon}{2}\cos(\omega_0 \tau)-\frac{\tau \epsilon^2}{4}\cos(2\omega_0\tau)$ and $\beta=\frac{\tau \epsilon^2}{4}>0$. 
The stability condition for each fixed point is given by
\begin{subequations}
\begin{align}
    &\frac{df_{\rm h}^1}{d\bar{R}}=\alpha-\frac{\epsilon C}{4}<0,\quad \left(\bar{R}^*=0\right), \label{eq:R0_stable_higher-order_control1}\\
    &\frac{df_{\rm h}^1}{d\bar{R}}=-2\left(\alpha+\beta
    -\frac{\epsilon C}{4}\right)<0,\quad \left(\bar{R}^*=1\right), 
    \label{eq:R1_stable_higher-order_control1}\\
    &\frac{df_{\rm h}^1}{d\bar{R}}=-2\beta\left(
    -\frac{\alpha}{\beta}+\frac{\epsilon C}{4\beta}\right)\left(-\frac{\alpha}{\beta}+\frac{\epsilon C}{4\beta}-1\right)<0,\notag\\
    &\hspace{36mm} \left(\bar{R}^*=\sqrt{-\frac{\alpha}{\beta}+\frac{\epsilon C}{4\beta}}\right).
    \label{eq:Rm_stable_higher-order_control1}
\end{align}
\label{eq:stability_higher_order_control1}
\end{subequations}
Thus, the stability of the fixed points $\bar{R}^*=0$ and $1$ 
changes depending on the  control parameter $C$, while the fixed point $\bar{R}^*=\sqrt{-\frac{\alpha}{\beta}+\frac{\epsilon C}{4\beta}}$ is always neutrally stable or unstable, i.e., no range of $C$ exists for which 
Eq.~\eqref{eq:Rm_stable_higher-order_control1} is satisfied (recall that $\bar{R}^* <1$).
Note that there exists a bistable region of $\bar{R}^*=0$ and  $\bar{R}^*=1$
in which Eqs.~\eqref{eq:R0_stable_higher-order_control1} and \eqref{eq:R1_stable_higher-order_control1} are simultaneously satisfied.

\subsubsection{Pairwise approximation}
From Eq.~\eqref{eq:averaging_Rp} and $D\left(\bar{R}\right)=C\bar{R}$,
we obtain the pairwise reduced equation as
\begin{align}
    \dot{\bar{R}}&=\bar{R}(1-\bar{R}^2)
    \left\{\frac{\epsilon}{2}\cos(\omega_0\tau)
    -\frac{\epsilon C}{4}
    \right\}\eqqcolon f_{\rm p}^1\left(\bar{R}\right).
    \label{eq:pairwise_control1}
\end{align}
From $f_{\rm p}^1\left(\bar{R}\right)=0$, we obtain that the fixed points are $\bar{R}^*=0$~and~$1$.
Note that only two fixed points can be found in the pairwise approximation.
The linear stability conditions are given by
\begin{subequations}
\begin{align}
&\frac{df_{\rm p}^1}{d\bar{R}}=\frac{\epsilon}{2}\left(\cos(\omega_0 \tau)-\frac{C}{2}\right)<0, \quad \left(\bar{R}^*=0\right),    \\
&\frac{df_{\rm p}^1}{d\bar{R}}=-\epsilon\left(\cos(\omega_0 \tau)-\frac{C}{2}\right)<0, \quad \left(\bar{R}^*=1\right).
\end{align}
\label{eq:stability_pairwise_control1}
\end{subequations}
\noindent
The stability of these fixed points depends on the control parameter $C$, and there is no bistable region of $\bar{R}^*=0$ and $\bar{R}^*=1$, in contrast to the case of 
the higher-order approximation. 

\begin{figure}[!tb]
    \centering
   \begin{minipage}{0.365\linewidth}
        \hspace{-9.1mm}
        \centering
        \includegraphics[width=\linewidth]{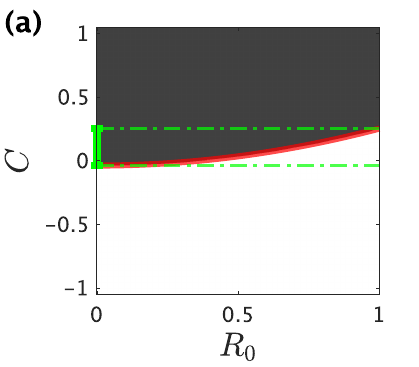}
    \end{minipage}%
    \begin{minipage}{0.365\linewidth}
        \centering
        \includegraphics[width=\linewidth]{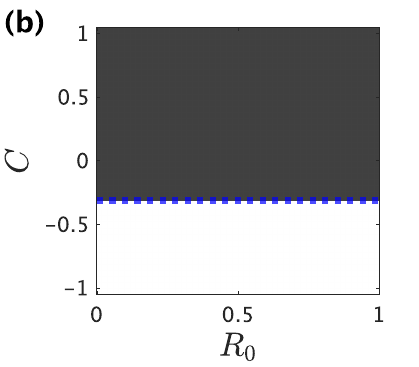}
    \end{minipage}
    \begin{minipage}{0.45\linewidth}
        \centering
        \vspace{2mm}
        \includegraphics[width=\linewidth]{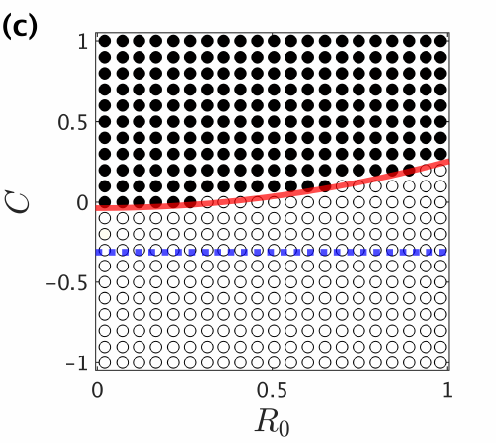}
    \end{minipage}
    \caption{
    Stability diagrams (a,b) and simulation results (c)
    under the control input $u=C R \cos\Psi$ with $\epsilon=0.1,~
    \tau=2.9,~\omega_0=\pi/2$ and $N=300$.
    (a): Higher-order reduced equation, (b): pairwise reduced equation.
    The black (white) region indicates that the solution with initial condition $R_0$ converges to $\bar{R}=0$ ($\bar{R}=1$).
    Note that the bistable region, which is indicated by the green highlighting along the $C$ axis, 
    appears only in the higher-order reduced equation in (a),
    where $\bar{R}$ converges to either $\bar{R}=0$ or $\bar{R}=1$ depending on $R_0$.
    (c): Simulation results of the original time-delayed Kuramoto model, Eq.~\eqref{eq:Kuramoto_time_delay_with_control}.
    Stability boundaries in (a) are plotted with red solid line and (b) with blue dotted line.
    The black (white) dots indicate that the numerical solution from the initial condition $R_0$ converges to $R=0$ ($R=1$).}
    \label{fig:CR0_tau29}
\end{figure}

\subsubsection{Stability diagram}
Figure~\ref{fig:CR0_tau29}~(a) shows the stability diagram 
of the higher-order reduced equation, Eq.~\eqref{eq:higher_order_control1}, whereas Figure~\ref{fig:CR0_tau29}~(b) shows that of the pairwise reduced equation, Eq.~\eqref{eq:pairwise_control1}.
Each figure shows the asymptotic value reached from the initial condition $R_0$ (horizontal axis) for each control parameter $C$ (vertical axis),
where black and white regions correspond to $\bar{R}=0$ and $\bar{R}=1$, respectively. 
Note that the fixed points $\bar{R}^*=0$ and $\bar{R}^*=1$ are bistable in a range of $C$ 
in the higher-order approximation, whereas no such bistability arises in the pairwise approximation.

\subsection{Case II: $u= CR(R-E)\cos\Psi$}

In this subsection, we consider the second type of feedback and
prove that any value of $\bar{R}$, rather than only $\bar{R}=0$ and $1$,
can be stabilized by adjusting the pair of control parameters $C$ and $E$.

\subsubsection{Higher-order approximation}

From Eq.~\eqref{eq:averaging_Rh} and $D\left(\bar{R}\right)=C\bar{R}\left(\bar{R}-E\right)$, the higher-order reduced equation is
\begin{widetext}
\begin{align}
    \dot{\bar{R}}&=\bar{R}(1-\bar{R}^2)
    \left\{
    \left(
    \frac{\epsilon}{2}\cos(\omega_0\tau)
    -\frac{\tau\epsilon^2}{4}\cos(2\omega_0\tau)
    +\frac{\tau\epsilon^2}{4}\bar{R}^2
    \right)-\frac{\epsilon C(\bar{R}-E)}{4}
    \right\}\eqqcolon f_{\rm h}^2\left(\bar{R}\right).
    \label{eq:higher_order_control2}
\end{align}
\end{widetext}
From $f_{\rm h}^2\left(\bar{R}\right)=0$, we can obtain four fixed points
\begin{align}
    \bar{R}^*=0,~1,~\bar{R}^*_{{\rm h},+},
    ~\mbox{and}~\bar{R}^*_{{\rm h},-},
\end{align}
where $\bar{R}^*_{{\rm h},\pm}=\frac{\frac{\epsilon C}{4\beta}\pm\sqrt{\left(\frac{\epsilon C}{4\beta}\right)^2-4\left(\frac{\alpha}{\beta}+\frac{\epsilon C}{4\beta}E\right)}}{2}$ and $\alpha$ and $\beta$ are defined as in Sec.~\ref{subsubsec:higher-order_control1}.

The stability condition for each fixed point is
{\small
\begin{subequations}
\begin{align}
    \frac{df_{\rm h}^2}{d\bar{R}}&=\alpha+\frac{\epsilon C}{4}E<0, \quad 
    \left(\bar{R}^*=0\right),\\
    \frac{df_{\rm h}^2}{d\bar{R}}&=-2\left(\alpha
    +\beta-\frac{\epsilon C}{4}\left(1-E\right)\right)<0, \quad \left(\bar{R}^*=1\right),\\
    \frac{df_{\rm h}^2}{d\bar{R}}&
    =\pm\beta\bar{R}^*_{{\rm h},\pm}
    \left(1-\left(\bar{R}^*_{{\rm h},\pm}\right)^2\right)
    \sqrt{\left(\frac{\epsilon C}{4\beta}\right)^2-4\left(\frac{\alpha}{\beta}+\frac{\epsilon C}{4\beta}E\right)}<0,\notag\\
    &\hspace{60mm}\left(\bar{R}^*=\bar{R}^*_{{\rm h},\pm}\right).
\end{align}
\label{eq:condition_higher-order_control2}
\end{subequations}}\noindent
The stability of $\bar{R}^*=0$ and $\bar{R}^*=1$ depends on the control parameters $C$ and $E$, whereas $\bar{R}^*_{{\rm h},-}$ ($\bar{R}^*_{{\rm h},+}$) is always linearly stable (unstable) whenever 
$0<\bar{R}^*_{{\rm h},\pm}<1$.
If there exists a region in which two of the conditions in Eq.~\eqref{eq:condition_higher-order_control2}
are satisfied simultaneously, the corresponding solutions are bistable.
\par

Note that the stable fixed point
\begin{align}
    \bar{R}^*_{{\rm h},-}=
    \frac{\frac{\epsilon C}{4\beta}-\sqrt{\left(\frac{\epsilon C}{4\beta}\right)^2-4\left(\frac{\alpha}{\beta}+\frac{\epsilon C}{4\beta}E\right)}}{2}.
    \label{eq:condition_higher-order_control2_Rref}
\end{align}
can  take any value $\bar{R}^*_{{\rm h},-} = R_{\rm ref}$ in $0<R_{\rm ref}<1$ by choosing the parameters.

\subsubsection{Pairwise approximation}
From Eq.~\eqref{eq:averaging_Rp} and $D\left(\bar{R}\right)=C\bar{R}(\bar{R}-E)$, 
the pairwise reduced equation is given by
\begin{align}
    \dot{\bar{R}}&=\bar{R}(1-\bar{R}^2)
    \left\{
    \frac{\epsilon}{2}\cos(\omega_0\tau)
    -\frac{\epsilon C(\bar{R}-E)}{4}
    \right\}
    \eqqcolon f_{\rm p}^2\left(\bar{R}\right).
    \label{eq:pairwise_control2}
\end{align}
From $f_{\rm p}^2\left(\bar{R}\right)=0$, we compute the following three fixed points
\begin{align}
    \bar{R}^*=0,~1,
    ~\mbox{and}
    ~\bar{R}^*_{\rm p},
\end{align}
where $\bar{R}^*_{\rm p}=\frac{2\cos(\omega_0 \tau)}{C}+E$.

The stability conditions for these fixed points are
\begin{subequations}
\begin{align}
    &\frac{df_{\rm p}^2}{d\bar{R}}=\frac{\epsilon C}{4}\left(\frac{2\cos(\omega_0 \tau)}{C}+E\right)<0,	\quad \left(\bar{R}^*=0\right),\\
    &\frac{df_{\rm p}^2}{d\bar{R}}=-\frac{\epsilon C}{2}\left(\frac{2\cos(\omega_0 \tau)}{C}+E-1\right)<0, \quad \left(\bar{R}^*=1\right),
    \label{eq:R1_stable_pairwise_control2}\\
    &\frac{df_{\rm p}^2}{d\bar{R}}=
    -\frac{\epsilon C}{4}\bar{R}^*_{\rm p}\left(1-\left(\bar{R}^*_{\rm p}\right)^2\right)<0. \quad \left(\bar{R}^*=R^*_{\rm p}\right).
\end{align}
\label{eq:condition_pairwise_control2}
\end{subequations}\noindent
If two of the conditions in Eq.~\eqref{eq:condition_pairwise_control2}
are satisfied simultaneously, the corresponding solutions are bistable.

The stability of $\bar{R}^*=0,~1$,~and~$\bar{R}^*_{\rm p}$ is determined by the pair of the control parameters $C$ and $E$.
The fixed point
\begin{align}
    \bar{R}^*_{\rm p}=\frac{2\cos(\omega_0 \tau)}{C}+E,\quad C>0.
    \label{eq:condition_pairwise_control2_Rref}
\end{align}
can take any value $\bar{R}^*_{\rm p}=R_{\rm ref}~(0<R_{\rm ref}<1)$ by choosing the parameters $C$ and $E$.

\subsubsection{Stability diagram}

\begin{figure}[tb]
    \centering
    \begin{minipage}{0.45\linewidth}
        \hspace{-9.1mm}
        \centering
        \includegraphics[width=\linewidth]{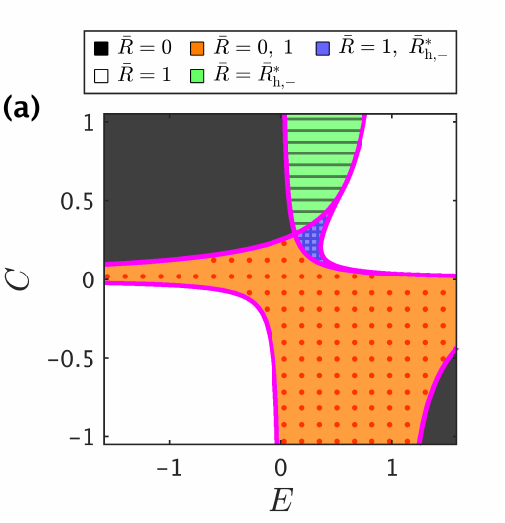}
    \end{minipage}%
    \begin{minipage}{0.45\linewidth}
        \centering
        \includegraphics[width=\linewidth]{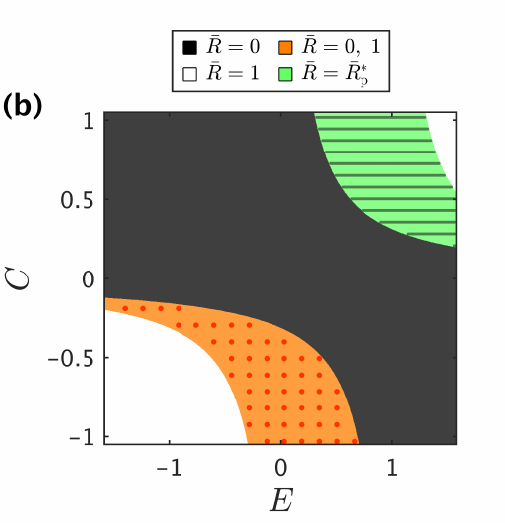}
    \end{minipage}
    \centering
    \begin{minipage}{0.495\linewidth}
        \centering
        \includegraphics[width=\linewidth]{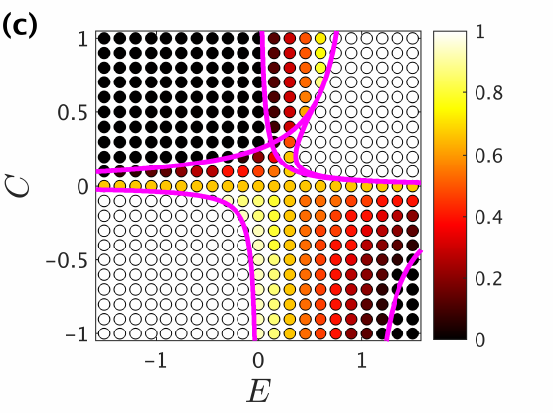}
    \end{minipage}%
    \begin{minipage}{0.495\linewidth}
        \centering
        \includegraphics[width=\linewidth]{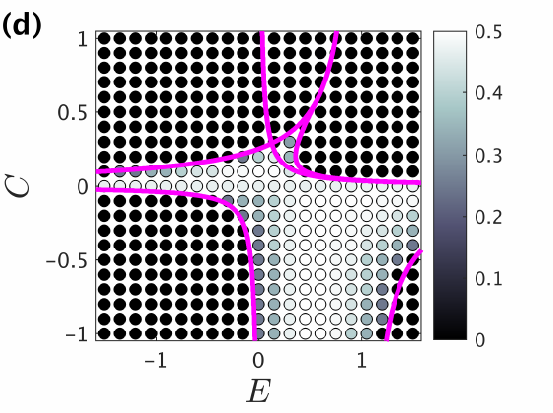}
    \end{minipage}
    \caption{
    Stability diagrams (a,b) and simulation results (c,d) under the control input $u= C R(R-E) \cos\Psi$ with $\epsilon=0.1,~
    \tau=2.9,~\omega_0=\pi/2$ and $N=300$.
    (a): Higher-order approximation, (b):  pairwise approximation.
    The black (white) region indicates that $\bar{R}=0$ ($\bar{R}=1$) is the unique stable state.
    The orange region corresponds to a bistable state of $\bar{R}=0$ and $\bar{R}=1$, and the green regions 
    show that $\bar{R}^*_{{\rm h},-}$ in (a) or $\bar{R}^*_{\rm p}$ 
    in (b) is the stable fixed point.
    The blue region indicates the bistability of $\bar{R}=1$ and $\bar{R}^*_{{\rm h},-}$ in (a).
	(c): Mean asymptotic values of $R$ and (d): standard deviation of $R$ obtained from numerical simulations of Eq.~\eqref{eq:Kuramoto_time_delay_with_control} averaged over $15$ initial conditions. 
    In (a), (c) and (d), the boundaries in panel (a) are overlaid in magenta.
    }
    \label{fig:CE_tau29}
\end{figure}

Figure~\ref{fig:CE_tau29} shows the stability diagrams for the higher-order reduced equation, Eq.~\eqref{eq:higher_order_control2} (panel (a)), and for the pairwise reduced equation,  Eq.~\eqref{eq:pairwise_control2} (panel (b)).
The stability regions determined by Eqs.~\eqref{eq:condition_higher-order_control2} and \eqref{eq:condition_pairwise_control2} are color-coded in the parameter plane of $C$ and $E$.
The black and white represent the parameter regions where $\bar{R}=0$ and $\bar{R}=1$ are stable, respectively.
The orange indicates bistability between $\bar{R}=0$ and $\bar{R}=1$.
In the green regions,
$\bar{R}=\bar{R}^*_{{\rm h},-}$ (higher-order approximation) or $\bar{R}=\bar{R}^*_{\rm p}$ (pairwise approximation) is stable, allowing the stable fixed point to be continuously tuned within $0 < \bar{R} < 1$ 
by adjusting the control parameters.
The desired state $\bar{R}=R_{\rm ref}$ is stabilized when the conditions in Eq.~\eqref{eq:condition_higher-order_control2_Rref} or \eqref{eq:condition_pairwise_control2_Rref}, represented as lines 
in each figure, are satisfied.
Finally, the blue regions in (a) represent the bistability of $\bar{R}=1$ 
and $\bar{R}^*_{{\rm h},-}$. 
Note that such bistability is not captured by the pairwise approximation,
which is evident from Eqs.~\eqref{eq:R1_stable_pairwise_control2}
and \eqref{eq:condition_pairwise_control2_Rref}.

\section{Numerical simulations}
\label{sec:numsim}

In this section, we perform numerical simulations of the time-delayed Kuramoto model Eq.~\eqref{eq:Kuramoto_time_delay_with_control} to demonstrate 
the validity of the stability regions, which are predicted by the approximate one-dimensional equation for $\bar{R}$ under the two types of feedback control, Eqs.~\eqref{eq:control1} and \eqref{eq:control2}.
To demonstrate the usefulness of the higher-order approximation for controlling the delayed system, we compare the predictive accuracy 
between the higher-order reduced equation and the pairwise reduced equation.

\subsection{Case I: $u=C R \cos\Psi$}

Figure~\ref{fig:CR0_tau29} (c) shows the asymptotic values of $R$ 
obtained by numerically simulating
Eq.~\eqref{eq:Kuramoto_time_delay_with_control} under the control input $u= C R \cos\Psi$,
along with the boundaries predicted by the higher-order and pairwise reduced
equations.
By adjusting the control parameter $C$, either $R=0$ (black dots) or $R=1$ (white dots) can be chosen as the unique stable state in the monostable region, where all initial conditions $R_0$ yield the same asymptotic state in this case. 
In the bistable region, the value of $C$ changes the relative sizes of the basins of 
attraction associated with $R=0$ and $R=1$.

It is clear that the simulation results in Fig.~\ref{fig:CR0_tau29} (c) better agree with the stability diagram predicted by the higher-order 
reduced equation in Fig.~\ref{fig:CR0_tau29} 
(a) than the one resulting from the pairwise reduced equation Fig.~\ref{fig:CR0_tau29} (b).
To quantitatively evaluate the predictive accuracies of the higher-order and pairwise approximations, Eqs.~\eqref{eq:higher_order_control1} and \eqref{eq:pairwise_control1}, we compute the errors between the simulation results of Eq.~\eqref{eq:Kuramoto_time_delay_with_control} 
in Fig.~\ref{fig:CR0_tau29} (c) and their predictions.
The Global root mean squared error (Global RMSE), computed over all simulation conditions, is defined as
\begin{equation}
    {\rm Global~RMSE}= \sqrt{\frac{1}{n}\sum_{j=1}^{n}\left(\bar{R}^*_j-R_j(t_f)\right)^2},
    \label{eq:RMSE}
\end{equation}
where each index $j$ corresponds to a particular parameter set, i.e., a couple $(R_0,C)$
in the case of $u=CR\cos\Psi$,
$n$ is the total number of couples,
$\bar{R}^{*}_{j}$ denotes the stable fixed point predicted by the reduced equation,
and $R_j(t_f)$ is the value obtained from the numerical simulation of 
Eq.~\eqref{eq:Kuramoto_time_delay_with_control} at the final time $t_f$.

The Global RMSE values are listed in Tab.~\ref{tab:CR0_RMSE} for two different values of the time delay, $\tau=2.9$ and $6.8$.
The higher-order reduced equation exhibits higher predictive accuracy than the pairwise reduced equation.
As expected, the Global RMSE increases with $\tau$, and even the higher-order approximation becomes less accurate for the larger delay,
especially around the boundary of different dynamical regimes
(See Appendix for the results with $\tau=6.8$).

\begin{table}[tb]
    \centering
    \begin{tabular}{|c|c|c|}
        \hline
        & ~higher-order~ & ~pairwise~       \\ \hline
        ~$\tau=2.9$~ & $4.8\times10^{-2}$ & $4.4\times10^{-1}$ \\ \hline
        ~$\tau=6.8$~ & $2.0\times10^{-1}$ & $6.5\times10^{-1}$ \\ \hline
    \end{tabular}
    \caption{Global RMSE for the higher-order and the pairwise reduced equations
    for $\tau=2.9,~6.8$ under the control input $u= C R \cos\Psi$ with $\epsilon=0.1,~\omega_0=\pi/2,~N=300$ and $n=21\times 21$~(21 values each for $R_0$ and $C$).
    }
    \label{tab:CR0_RMSE}
\end{table}

\subsection{Case II: $u=CR(R-E)\cos\Psi$}

In Figs.~\ref{fig:CE_tau29} (c) and (d), we show the simulation results of Eq.~\eqref{eq:Kuramoto_time_delay_with_control} with the control input $u=CR(R-E)\cos\Psi$.
Each dot shows the mean asymptotic value of $R$ over $15$ random initial conditions $R_0$ in (c), and the corresponding standard deviation in (d).
In panel (c), white dots (black dots) indicate regions where $R=0$ $(R=1)$ is the only stable state. The remaining dots are colored according to the mean asymptotic value of $R$, which lies between $0$ and $1$, indicating either the existence of a monostable intermediate fixed point or bistability.
In panel (d), black dots indicate monostable regions, whereas the other dots indicate bistable regions,
allowing us to distinguish between monostable intermediate states and bistable states.

We now compare the predictive accuracy of the higher-order and the pairwise reduced equations, given by Eqs.~\eqref{eq:higher_order_control2} and \eqref{eq:pairwise_control2}, respectively.
The Global RMSE values Eq.~\eqref{eq:RMSE}, now defined for the triplet  $(R_0,C,E)$, are listed in Tab.~\ref{tab:CE_RMSE} for $\tau=2.9$ and $6.8$.
In both cases, the higher-order reduced equation exhibits higher predictive accuracy compared with the pairwise reduced equation, although the prediction errors are relatively larger than those for the control input $u= C R\cos\Psi$ due to the more complex bifurcation structures.
(See Appendix~\ref{app:addnum} for the results with $\tau=6.8$ and error distributions).

\begin{table}[!tb]
    \centering
    \begin{tabular}{|c|c|c|}
        \hline
        & ~higher-order~ & ~pairwise~       \\ \hline
        ~$\tau=2.9$~ & $1.1\times10^{-1}$ & $5.3\times10^{-1}$ \\ \hline
        ~$\tau=6.8$~ & $2.8\times10^{-1}$ & $7.2\times10^{-1}$ \\ \hline
    \end{tabular}
    \caption{Global RMSE for the higher-order and the pairwise reduced equations
    for $\tau=2.9,~6.8$ under the control input $u=C R(R-E) \cos\Psi$ with $\epsilon=0.1,~\omega_0=\pi/2,~N=300$ and 
    $n=15\times 21\times 21$~(15 values for $R_0$, and 21 values each for $C$ and $E$).}
    \label{tab:CE_RMSE}
\end{table}

\subsection{Numerical demonstration of the stabilization of $R_{\rm ref}$}

\begin{figure}[tb]
    \centering
    \begin{minipage}{0.495\linewidth}
        \centering
        \includegraphics[width=\linewidth]{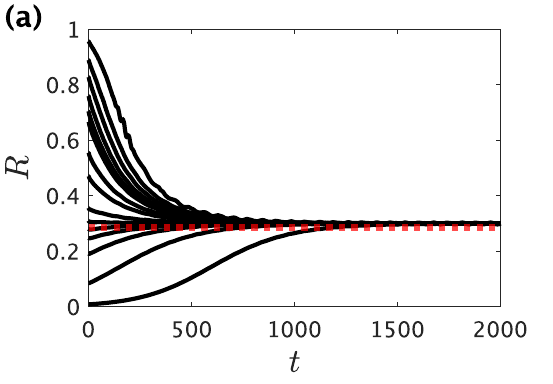}
    \end{minipage}%
    \begin{minipage}{0.47\linewidth}
        \vspace{1.5mm}
        \centering
        \includegraphics[width=\linewidth]{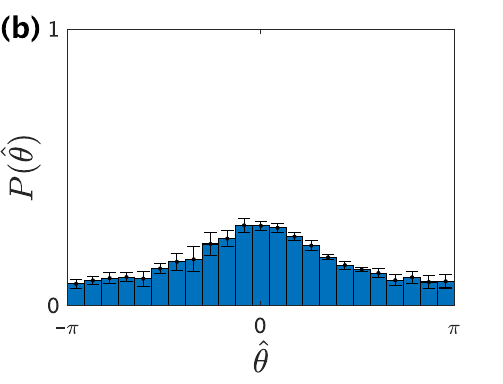}
    \end{minipage}
    \begin{minipage}{0.495\linewidth}
        \centering
        \includegraphics[width=\linewidth]{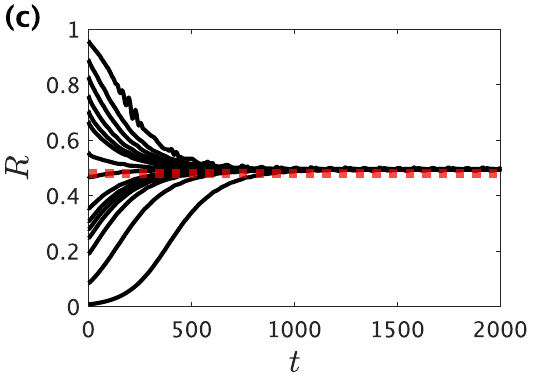}
    \end{minipage}%
    \begin{minipage}{0.47\linewidth}
        \vspace{1.5mm}
        \centering
        \includegraphics[width=\linewidth]{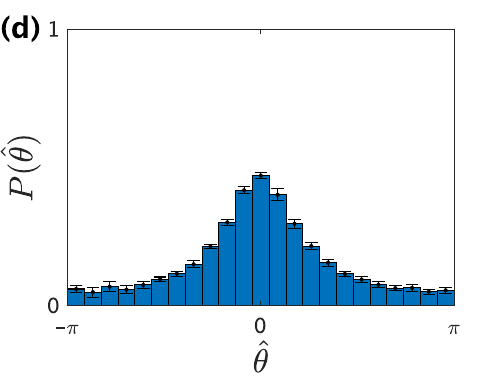}
    \end{minipage}
    \begin{minipage}{0.495\linewidth}
        \centering
        \includegraphics[width=\linewidth]{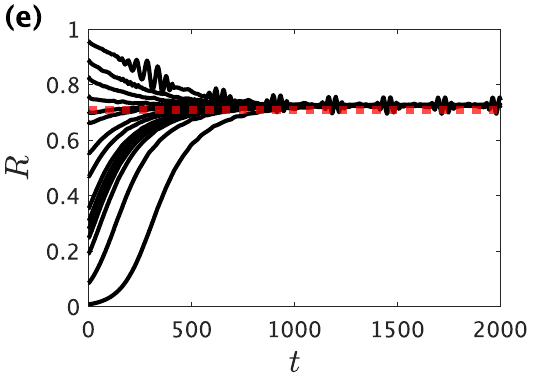}
    \end{minipage}%
    \begin{minipage}{0.47\linewidth}
        \vspace{1.5mm}
        \centering
        \includegraphics[width=\linewidth]{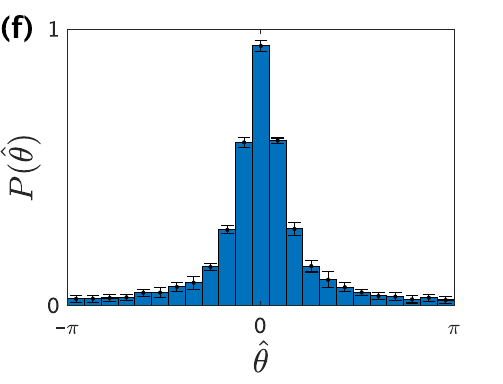}
    \end{minipage}
    \caption{Stabilization of the intermediate values of $R_{\rm ref}$ based on
    Eq.~\eqref{eq:condition_higher-order_control2_Rref} 
    with $\epsilon=0.1,~\omega_0=\pi/2,~\tau=2.9,~N=300$.
    In (a), (c), and (e), black solid lines show simulation results of the time-delayed Kuramoto model, given by Eq.~\eqref{eq:Kuramoto_time_delay_with_control},
    and red dotted lines show the corresponding $R_{\rm ref}$.
    Panels (b), (d), and (f) show the phase distributions $P(\hat{\theta})$,
    where $\hat{\theta}=\theta-\Psi$,
    at $t_f=2000$ averaged over $15$ random initial conditions.
	The parameter sets are as follows:
    (a) and (b): $(R_{\rm ref},~C,~E)=(0.29,~0.9,~0.3)$,
    (c) and (d): $(R_{\rm ref},~C,~E)=(0.48,~1.0,~0.45)$,
    and (e) and (f): $(R_{\rm ref},~C,~E)=(0.71,~1.0,~0.6)$.}
    \label{fig:control_results_CE_Rref}
\end{figure}

Finally, we show that an intermediate value $R_{\rm ref}$ can be stabilized by using the nonlinear mean-field feedback control
$u=CR(R-E)\cos\Psi$ according to Eq.~\eqref{eq:condition_higher-order_control2_Rref}, derived from the higher-order reduced equation, and further demonstrate the existence of a bistable regime in which both $R=1$ and $R=R_{\rm ref}$ are stable.
Figure \ref{fig:control_results_CE_Rref} shows simulation results of Eq.~\eqref{eq:Kuramoto_time_delay_with_control} from $15$ random initial conditions, where the control parameters are chosen from the green region in Fig.~\ref{fig:CE_tau29} (a).
Panels (a), (c), and (e) show the time series of $R$ in black with corresponding reference value $R=R_{\rm ref}$ in red. 
Panels (b), (d), and (f) show the phase distributions $P(\hat{\theta})$ of the oscillator phases relative to the collective phase,
 $\hat{\theta}=\theta-\Psi$,
at the final time $t_f$ of the simulations averaged over $15$ random initial conditions, together with the corresponding standard deviations.
All trajectories converge close to $R=R_{\rm ref}$, 
indicating successful stabilization at 
the desired value of the order parameter and the phase distributions exhibit different 
degrees of spread depending on $R_{\rm ref}$.

\begin{figure}[tb]
    \centering
    \begin{minipage}{0.495\linewidth}
        \centering
        \includegraphics[width=\linewidth]{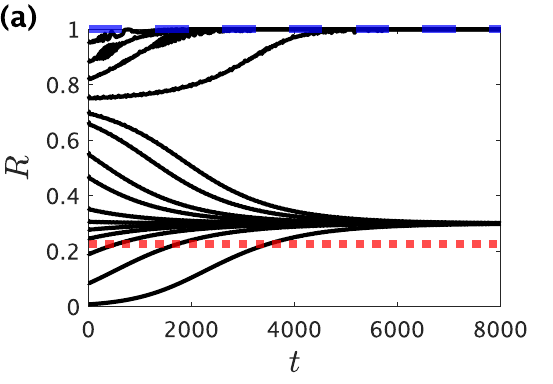}
    \end{minipage}%
    \begin{minipage}{0.47\linewidth}
        \vspace{1.5mm}
        \centering
        \includegraphics[width=\linewidth]{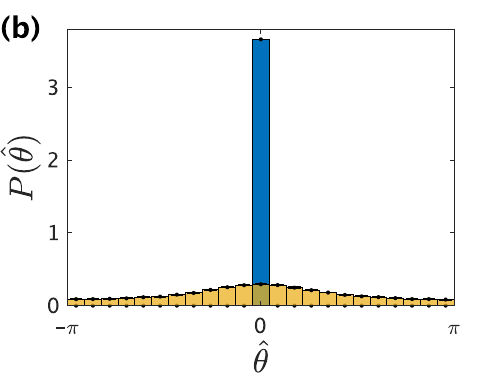}
    \end{minipage}
    \caption{Bistability of the fully synchronized state and an intermediate synchronized state predicted by the higher-order reduced equation with $\epsilon=0.1$, $\omega_0=\pi/2$, $\tau=2.9$, and $N=300$ for $C=0.3,~E=0.3$ and the corresponding intermediate value
    $R=0.23$.
    In panel (a), black solid lines show simulation results of the time-delayed Kuramoto model given by Eq.~\eqref{eq:Kuramoto_time_delay_with_control}, while the red dotted (blue dashed) line indicates $R=0.23$ ($R=1$).
    Panel (b) shows the phase distributions $P(\hat{\theta})$, where $\hat{\theta}=\theta-\Psi$, at $t_f=8000$ averaged over $15$ random initial conditions.
    The phase distributions are computed separately for trajectories converging close to $R=0.23$ in yellow and those converging to $R=1$ in blue.}
    \label{fig:control_results_CE_Rref_bistability}
\end{figure}

Next, Figure \ref{fig:control_results_CE_Rref_bistability} represents simulation results 
in the bistable regime, where the control parameters are chosen from the blue region in Fig.~\ref{fig:CE_tau29} (a).
Panel (a) shows the time series of $R$ (black) with corresponding bistable reference values $R=R_{\rm ref}$ (red)
and $R=1$ (blue). 
Panel (b) shows the phase distributions and the corresponding standard deviations, where the $15$ trajectories are classified according to whether they converge to $R=R_{\rm ref}$ (yellow) or to $R=1$ (blue).
As predicted by the higher-order reduced equation, bistability of the fully synchronized state and an intermediate synchronized state is realized. 
However, the asymptotic value deviates noticeably from the predicted value $R_{\rm ref}$; more accurate predictions are attained for smaller values of $\epsilon\tau$.

\section{Conclusion}
\label{sec:conc}

In this work, we presented a framework for controlling the time-delayed Kuramoto model by using a delay-free reduced model obtained through a higher-order approximation.
We derived a one-dimensional reduced equation for the order parameter from the time-delayed Kuramoto model by combining the higher-order approximation, the OA ansatz, and the method of averaging.
We validated the proposed framework by comparing numerical simulations of the original time-delayed Kuramoto model with the predicted order-parameter dynamics of the reduced equation under two types of control, the linear and nonlinear mean-field feedback control. For comparison, we also performed a pairwise approximation and demonstrated that
the higher-order reduced equation exhibited higher predictive accuracy than the pairwise approximation. 
We demonstrated that a variety of desired states, including the incoherent state, fully synchronized state, bistable states, and intermediate synchronized states, were successfully 
realized in the time-delayed Kuramoto model using the nonlinear 
mean-field feedback control with the control parameters determined from the higher-order reduced model.
Therefore, the proposed higher-order approximation provides a delay-free reduced model that retains the predictive accuracy of the original time-delayed system while facilitating faster numerical simulations and the analytical design of the feedback control.

We expect that the proposed delay-free control framework for the time-delayed Kuramoto model can be extended to other classes of time-delayed dynamical networks, where delay effects are generally more difficult to analyze and control.

\section*{Acknowledgments}
We thank B. Rink for productive discussion.
N.F. was supported by JST SPRING, Japan Grant Number JPMJSP2180 and
the Science Tokyo Support Program for Doctoral Students,
funded by the Universities for International Research Excellence.
H.N. acknowledges JSPS KAKENHI 25H01468 and 25K03081 for financial support.
M.L. is a Postdoctoral Researcher of the Fonds de la Recherche Scientifique–FNRS.

\section*{Data Availability}
The data that supports the findings of this study are available within the article.

\appendix

\section{Derivation of the Averaged One-Dimensional Equation}
\label{app:derive_one}
We here show the details of the derivation of the averaged one-dimensional equation, 
given by Eq.~\eqref{eq:averaging_Rh}, from
Eqs.~\eqref{eq:OA_with_control_R_Psi_2},\eqref{eq:transformation} and \eqref{eq:averaging}.

By taking the time derivative of Eq.~\eqref{eq:transformation}, we obtain
{\small
\begin{subequations}
\begin{align}
    \dot{R}&=\dot{\bar{R}}+\epsilon \left\{\frac{\partial g^{(1)}}{\partial \bar{R}}\dot{\bar{R}}+\frac{\partial g^{(1)}}{\partial \bar{\Psi}}\dot{\bar{\Psi}}\right\}+\epsilon^2\left\{\frac{\partial g^{(2)}}{\partial \bar{R}}\dot{\bar{R}}+\frac{\partial g^{(2)}}{\partial \bar{\Psi}}\dot{\bar{\Psi}}\right\},\\
    \dot{\Psi}&=\dot{\bar{\Psi}}+\epsilon \left\{\frac{\partial l^{(1)}}{\partial \bar{R}}\dot{\bar{R}}+\frac{\partial l^{(1)}}{\partial \bar{\Psi}}\dot{\bar{\Psi}}\right\}+\epsilon^2\left\{\frac{\partial l^{(2)}}{\partial \bar{R}}\dot{\bar{R}}+\frac{\partial l^{(2)}}{\partial \bar{\Psi}}\dot{\bar{\Psi}}\right\}.
\end{align}
\label{eq:dt_transformation}
\end{subequations}
}\noindent
By substituting Eqs.~\eqref{eq:OA_with_control_R_Psi_2} and \eqref{eq:averaging} into Eq.~\eqref{eq:dt_transformation}
and comparing the coefficients of the same powers of $\epsilon$, we obtain
\begin{widetext}
\begin{subequations}
\begin{align}
&h^{(1)}(\bar R)
+\omega_0\frac{\partial g^{(1)}}{\partial\bar\Psi}=
F^{(1)}(\bar R,\bar\Psi),
\label{eq:order1_R}\\
&q^{(1)}(\bar R)
+\omega_0\frac{\partial l^{(1)}}{\partial\bar\Psi}=
G^{(1)}(\bar R,\bar\Psi),
\label{eq:order1_Psi}\\
&h^{(2)}(\bar R)
+\omega_0\frac{\partial g^{(2)}}{\partial\bar\Psi}=
\left.\frac{\partial F^{(1)}}{\partial R}\right|_{\bar R,\bar\Psi}g^{(1)}\left(\bar{R}, \bar{\Psi}\right)
+\left.\frac{\partial F^{(1)}}{\partial \Psi}\right|_{\bar R,\bar\Psi}l^{(1)}\left(\bar{R}, \bar{\Psi}\right)
+F^{(2)}(\bar R)-\frac{\partial g^{(1)}}{\partial\bar R}h^{(1)}(\bar R)
-\frac{\partial g^{(1)}}{\partial\bar\Psi}q^{(1)}(\bar R),
\label{eq:order2_R}\\
&q^{(2)}(\bar R)
+\omega_0\frac{\partial l^{(2)}}{\partial\bar\Psi}=
\left.\frac{\partial G^{(1)}}{\partial R}\right|_{\bar R,\bar\Psi}g^{(1)}\left(\bar{R}, \bar{\Psi}\right)
+\left.\frac{\partial G^{(1)}}{\partial \Psi}\right|_{\bar R,\bar\Psi}l^{(1)}\left(\bar{R}, \bar{\Psi}\right)
+G^{(2)}(\bar R)-\frac{\partial l^{(1)}}{\partial\bar R}h^{(1)}(\bar R)
-\frac{\partial l^{(1)}}{\partial\bar\Psi}q^{(1)}(\bar R).
\label{eq:order2_Psi}
\end{align}
\end{subequations}
\end{widetext}
From Eqs.~\eqref{eq:order1_R} and \eqref{eq:order1_Psi},
we derive
\begin{subequations}
\begin{align}
    &g^{(1)}\left(\bar{R}, \bar{\Psi}\right)
    =\frac{1}{\omega_0}\int_0^{\bar{\Psi}}\left\{F^{(1)}\left(\bar{R}, \bar{\Psi}'\right)-h^{(1)}\left(\bar{R}\right)\right\}d\bar{\Psi}',\\
    &l^{(1)}\left(\bar{R}, \bar{\Psi}\right)= \frac{1}{\omega_0}\int_0^{\bar{\Psi}}\left\{G^{(1)}\left(\bar{R}, \bar{\Psi}'\right)- q^{(1)}\left(\bar{R}\right)\right\}d\bar{\Psi}'.
\end{align}
\label{eq:first_order_gl}
\end{subequations}\noindent
In the method of averaging~\cite{verhulst2007averaging}, $h^{(1)}$ and $q^{(1)}$ are chosen as
\begin{subequations}
\begin{align}
    &h^{(1)}\left(\bar{R}\right)
    =\frac{1}{2\pi}\int_0^{2\pi}F^{(1)}\left(\bar{R}, \bar{\Psi}'\right)d\bar{\Psi}',\\
    &q^{(1)}\left(\bar{R}\right)= \frac{1}{2\pi}\int_0^{2\pi}G^{(1)}\left(\bar{R}, \bar{\Psi}'\right)d\bar{\Psi}'.
\end{align}
\label{eq:first_order_hq}
\end{subequations}
From Eqs.~\eqref{eq:order2_R} and \eqref{eq:order2_Psi}, we obtain
\begin{widetext}
{\small
\begin{subequations}
\begin{align}
    &g^{(2)}\left(\bar{R}, \bar{\Psi}\right)=\frac{1}{\omega_0}\int_0^{\bar{\Psi}}\left\{\left.\frac{\partial  F^{(1)}}{\partial R}\right|_{\bar{R}, \bar{\Psi}'}g^{(1)}\left(\bar{R}, \bar{\Psi}'\right)+\left.\frac{\partial  F^{(1)}}{\partial \Psi}\right|_{\bar{R}, \bar{\Psi}'}l^{(1)}\left(\bar{R}, \bar{\Psi}'\right)+F^{(2)}\left(\bar{R}\right)-\frac{\partial g^{(1)}}{\partial \bar{R}}h^{(1)}\left(\bar{R}\right)-\frac{\partial g^{(1)}}{\partial \bar{\Psi}'}q^{(1)}\left(\bar{R}\right)-h^{(2)}\left(\bar{R}\right)
    \right\}d{\bar{\Psi}}',\\
    &l^{(2)}\left(\bar{R}, \bar{\Psi}\right)=\frac{1}{\omega_0}\int_0^{\bar{\Psi}}\left\{\left.\frac{\partial  G^{(1)}}{\partial R}\right|_{\bar{R}, \bar{\Psi}'}g^{(1)}\left(\bar{R}, \bar{\Psi}'\right)+\left.\frac{\partial  G^{(1)}}{\partial \Psi}\right|_{\bar{R}, \bar{\Psi}'}l^{(1)}\left(\bar{R}, \bar{\Psi}'\right)+G^{(2)}\left(\bar{R}\right)-\frac{\partial l^{(1)}}{\partial \bar{R}} h^{(1)}\left(\bar{R}\right)-\frac{\partial l^{(1)}}{\partial \bar{\Psi}}q^{(1)}\left(\bar{R}\right)-q^{(2)}\left(\bar{R}\right)\right\}d{\bar{\Psi}}',
\end{align}
\label{eq:second_order_gl}
\end{subequations}
}\noindent
where $h^{(2)},~q^{(2)}$ are chosen as 
{\small
\begin{subequations}
\begin{align}
    &h^{(2)}\left(\bar{R}\right)
    =\frac{1}{2\pi}\int_0^{2\pi}\left\{\left.\frac{\partial  F^{(1)}}{\partial R}\right|_{\bar{R}, \bar{\Psi}'}g^{(1)}\left(\bar{R}, \bar{\Psi}'\right)+\left.\frac{\partial  F^{(1)}}{\partial \Psi}\right|_{\bar{R}, \bar{\Psi}'}l^{(1)}\left(\bar{R}, \bar{\Psi}'\right)+F^{(2)}\left(\bar{R}\right)-\frac{\partial g^{(1)}}{\partial \bar{R}}h^{(1)}\left(\bar{R}\right)-\frac{\partial g^{(1)}}{\partial \bar{\Psi}'}q^{(1)}\left(\bar{R}\right)
    \right\}d\bar{\Psi}',\\
    &q^{(2)}\left(\bar{R}\right)= \frac{1}{2\pi}\int_0^{2\pi}\left\{\left.\frac{\partial  G^{(1)}}{\partial R}\right|_{\bar{R}, \bar{\Psi}'}g^{(1)}\left(\bar{R}, \bar{\Psi}'\right)+\left.\frac{\partial  G^{(1)}}{\partial \Psi}\right|_{\bar{R}, \bar{\Psi}'}l^{(1)}\left(\bar{R}, \bar{\Psi}'\right)+G^{(2)}\left(\bar{R}\right)-\frac{\partial l^{(1)}}{\partial \bar{R}} h^{(1)}\left(\bar{R}\right)-\frac{\partial l^{(1)}}{\partial \bar{\Psi}}q^{(1)}\left(\bar{R}\right)\right\}d\bar{\Psi}'.
\end{align}
\label{eq:sacond_order_hq}
\end{subequations}
}\noindent\hspace{-2mm}
After solving Eqs.~\eqref{eq:first_order_gl}--\eqref{eq:sacond_order_hq} sequentially, we obtain 
Eq.~\eqref{eq:averaging_Rh}.
\end{widetext}

\section{Additional numerical results}
\label{app:addnum}

In this Appendix, we present additional numerical results for a larger time delay $\tau$ and the distribution of the prediction errors.

\subsection{Results for larger time delay}

We here present the results for the case with $\epsilon = 0.1$ and $\tau = 6.8$ for the two types of feedback control, where the larger value of $\epsilon\tau$ is expected to reduce the accuracy of the approximation, and compare the results with those obtained for $\epsilon = 0.1$ and 
$\tau = 2.9$.

\begin{figure}[!b]
    \centering
   \begin{minipage}{0.365\linewidth}
        \hspace{-9.1mm}
        \centering
        \includegraphics[width=\linewidth]{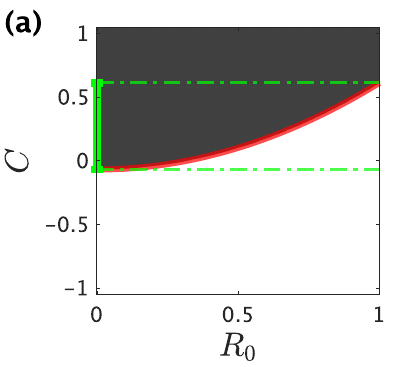}
    \end{minipage}%
    \begin{minipage}{0.365\linewidth}
        \centering
        \includegraphics[width=\linewidth]{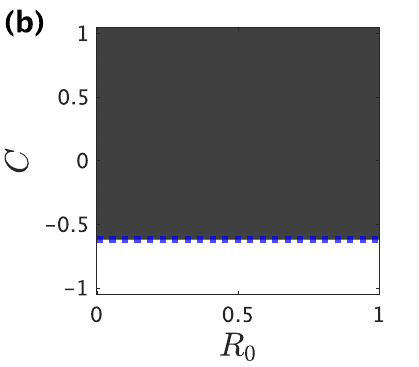}
    \end{minipage}
    \begin{minipage}{0.45\linewidth}
        \vspace{2mm}
        \centering
        \includegraphics[width=\linewidth]{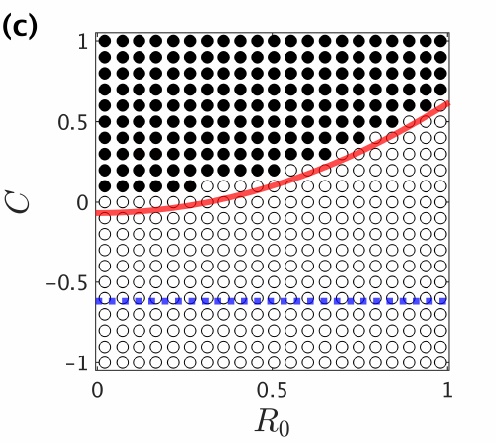}
    \end{minipage}
    \caption{
    Stability diagrams and simulation results
    under the control input $u=C R \cos\Psi$.
    In the same manner as Fig.~\ref{fig:CR0_tau29}, but for
    $\epsilon=0.1,~
    \tau=6.8,~\omega_0=\pi/2$ and $N=300$.}
    \label{fig:CR0_tau68}
\end{figure}

\begin{figure}[tb]
    \centering
    \begin{minipage}{0.45\linewidth}
        \hspace{-9.1mm}
        \centering
        \includegraphics[width=\linewidth]{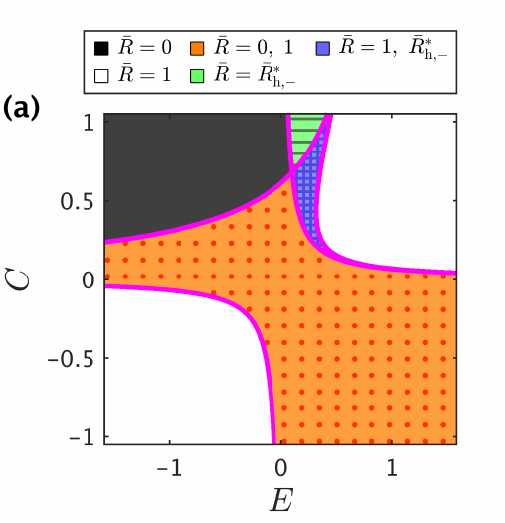}
    \end{minipage}%
    \begin{minipage}{0.45\linewidth}
        \centering
        \includegraphics[width=\linewidth]{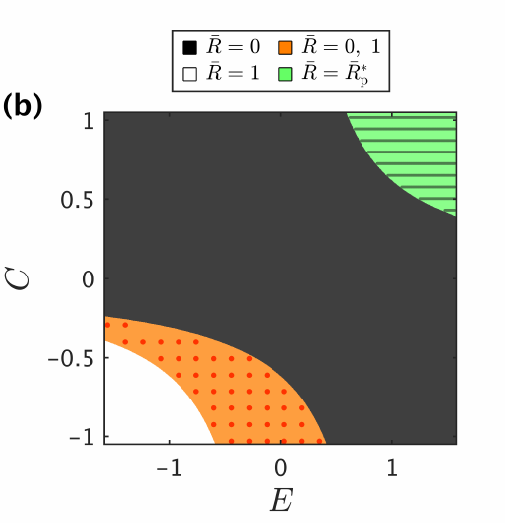}
    \end{minipage}
    \centering
    \begin{minipage}{0.495\linewidth}
        \centering
        \includegraphics[width=\linewidth]{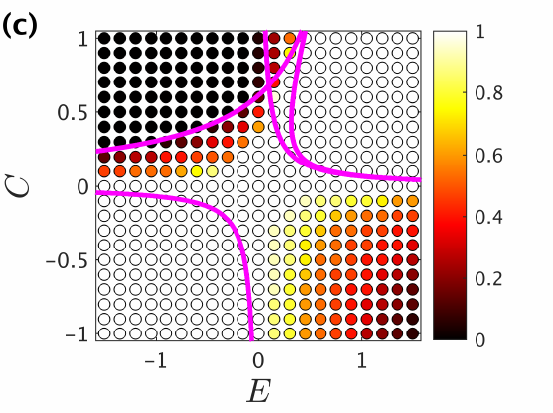}
    \end{minipage}%
    \begin{minipage}{0.495\linewidth}
        \centering
        \includegraphics[width=\linewidth]{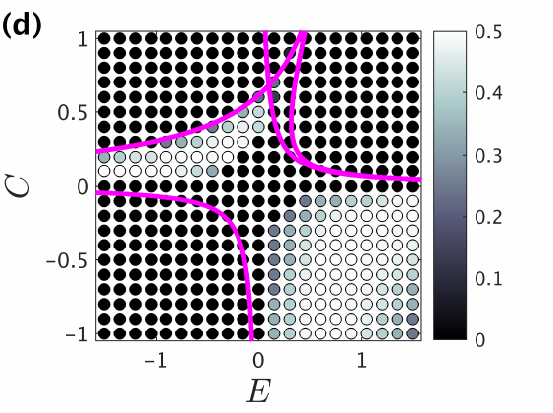}
    \end{minipage}
    \caption{
    Stability diagrams and simulation results
    under the control input $u=C R(R-E)\cos\Psi$.
    In the same manner as Fig.~\ref{fig:CE_tau29}, but for
    $\epsilon=0.1,~
    \tau=6.8,~\omega_0=\pi/2$ and $N=300$.}
    \label{fig:CE_tau68}
\end{figure}

Figures \ref{fig:CR0_tau68}~(a) and (b) show 
the stability diagrams obtained from Eqs.~\eqref{eq:higher_order_control1} and
\eqref{eq:pairwise_control1}
for $u=CR\cos\Psi$, respectively, and
Fig.~\ref{fig:CR0_tau68}~(c) shows simulation results
of Eq.~\eqref{eq:Kuramoto_time_delay_with_control}.
The results for the control input $u=CR(R-E)\cos\Psi$ are shown in
Fig.~\ref{fig:CE_tau68}.
Panels (a) and (b) present the stability diagrams obtained from
Eqs.~\eqref{eq:higher_order_control2} and
\eqref{eq:pairwise_control2}, respectively,
while panels (c) and (d) show simulation results of
Eq.~\eqref{eq:Kuramoto_time_delay_with_control}.

To quantify the predictive accuracy of the stability diagrams
for the simulation results, the Global RMSE values, given by 
Eq.~\eqref{eq:RMSE}, are summarized in
Tables~\ref{tab:CR0_RMSE} and \ref{tab:CE_RMSE}.
In addition, the distributions of the prediction errors for the case of $u=CR(R-E)\cos\Psi$
are presented in Sec.~\ref{sec:error_distribution}.
As expected, the higher-order approximation yields more accurate predictions than the pairwise approximation, although even the higher-order reduced equation shows less accurate as $\epsilon\tau$ increases.
In particular, the boundary of the predicted regions from the higher-order reduced equation
in Fig.~\ref{fig:CR0_tau68} (a) is noticeably shifted from the boundary in Fig.~\ref{fig:CR0_tau68} (c). 
In Fig.~\ref{fig:CE_tau68}, the higher-order reduced equation predicts a bistable region between $R=0$ and $R=1$ (the orange region) in (a), whereas, in part of this region, the simulations 
exhibit a monostable fully synchronized state at $R=1$ in (c) and (d).

\subsection{Distribution of Prediction Errors}
\label{sec:error_distribution}

We here present the distribution of the prediction errors
to provide a more detailed assessment of the discrepancy between
the predictions of the one-dimensional reduced equation and the simulation results of the original time-delayed Kuramoto model.
For the case $u=CR\cos\Psi$, the discrepancy is already evident from Figs.~\ref{fig:CR0_tau29} (c)
and \ref{fig:CR0_tau68} (c) because the stabilized states are $R=0$ or $R=1$. 
Therefore, we here focus on the results for $u=CR(R-E)\cos\Psi$.

\begin{figure}[tb]
    \centering
    \begin{minipage}{0.495\linewidth}
        \centering
        \includegraphics[width=\linewidth]{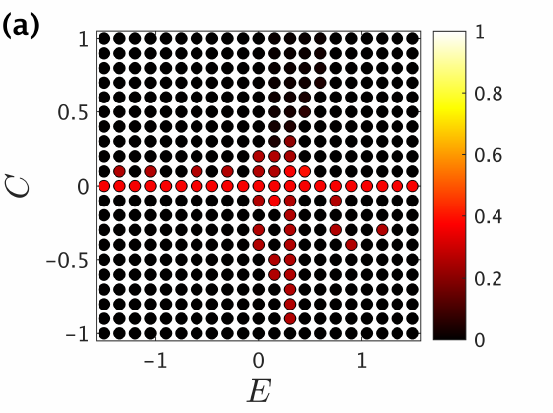}
    \end{minipage}%
    \begin{minipage}{0.495\linewidth}
        \centering
        \includegraphics[width=\linewidth]{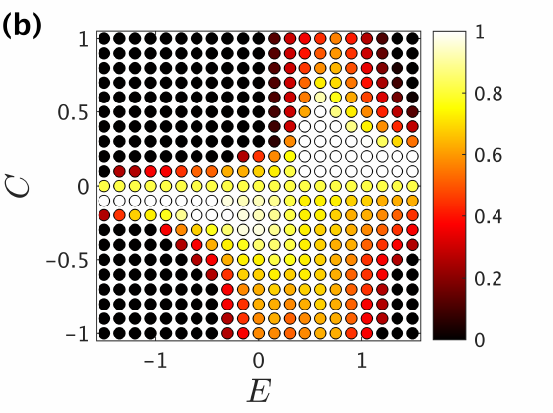}
    \end{minipage}
    \centering
    \begin{minipage}{0.495\linewidth}
        \centering
        \includegraphics[width=\linewidth]{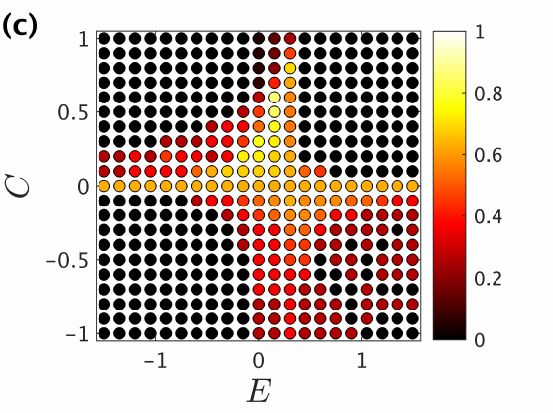}
    \end{minipage}%
    \begin{minipage}{0.495\linewidth}
        \centering
        \includegraphics[width=\linewidth]{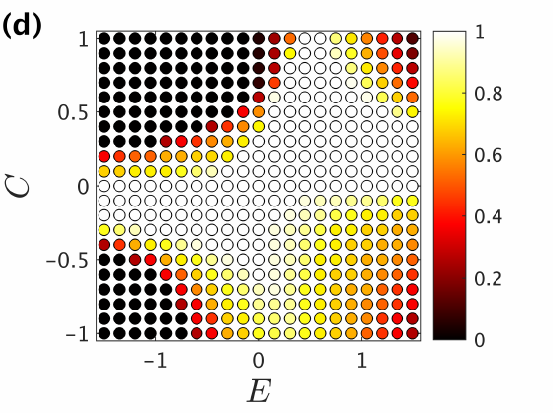}
    \end{minipage}
    \caption{
    Distribution of RMSE over $15$ random initial conditions for the higher-order reduced equation in (a) and (c), 
    and that of the pairwise reduced equation in (b) and (d), 
    $u=C R(R-E) \cos\Psi$ with $\epsilon=0.1,~\omega_0=\pi/2,~N=300$.
    Panels (a) and (b) show the case of $\tau=2.9$, whereas panels (c) and (d) show the case $\tau=6.8$.}
    \label{fig:CE_error}
\end{figure}

Figure \ref{fig:CE_error} shows the distribution of RMSE over $15$ random initial conditions.
Panels (a) and (c) show the errors of the higher-order approximation
and panels (b) and (d) show the errors for the pairwise approximation for 
the case of $\tau=2.9$ and $6.8$, respectively.
As expected, the higher-order approximation yields  
fewer regions of large errors than the pairwise approximation.
However, even with the higher-order approximation, the prediction errors increase for larger $\tau$, especially in regions exhibiting bistability and stable intermediate values of $R$.


\end{document}